\documentclass[]{aa}

\newcommand{\e}{\times10^}

\usepackage{lscape}
\usepackage{graphicx}
\usepackage{natbib}
\usepackage{longtable}
\usepackage{supertabular}
\bibpunct{(}{)}{;}{a}{}{,}
\usepackage{txfonts}
\begin{document}

   \title{The onset of star formation in primordial haloes}
   \titlerunning{The onset of star formation}
   \authorrunning{Umberto Maio et al.}

   \author{Umberto  Maio      \inst{1,2},
          Benedetta Ciardi    \inst{1},
          Naoki     Yoshida   \inst{3},
          Klaus     Dolag     \inst{1}
          \and
          Luca      Tornatore \inst{4}
          }

   \offprints{Umberto Maio\\e-mail: maio@mpa-garching.mpg.de, umaio@mpe.mpg.de}

   \institute{Max-Planck-Institut f\"ur Astrophysik, 
     Karl-Schwarzschild-Stra{\ss}e 1, 85748 Garching bei M\"unchen, Germany
     \and
     Max-Planck-Institut f\"ur extraterrestrische Physik, 
     Giessenbachstra{\ss}e 1, 85748 Garching bei M\"unchen, Germany
     \and
     IPMU, U-Tokyo
     5-1-5 Kashiwanoha, Kashiwa, Chiba 277-8568, Japan
     \and
     INAF -- Osservatorio astronomico di Trieste,
     Via Tiepolo 11, 34143 Trieste, Italy.
   }

   \date{Received ...; ...}

 
  \abstract
      {Star formation remains an unsolved problem in astrophysics.
        Numerical studies of large-scale structure simulations cannot
        resolve the process and their approach usually
        assumes that only gas denser than a typical
        threshold can host and form stars.}
   {We investigate the onset of cosmological star formation and
     compare several very-high-resolution, three-dimensional,
     N-body/SPH simulations that include non-equilibrium,
     atomic and molecular chemistry, star formation prescriptions, and
     feedback effects.}
   {We study how primordial star formation
     depends on gas density threshold,
     cosmological parameters, and initial set-ups.}
   {For mean-density initial conditions, we find that standard
     low-density star-formation threshold ($0.2\,h^2\,\rm cm^{-3}$)
     models predict the onset of star formation at $z\sim~25-31$,
     depending on the adopted cosmology.
     In these models, stars are formed automatically when the gas
     density increases above the adopted threshold, regardless of the time
     between the moment when the threshold is reached and the
     effective runaway collapse. While this is a reasonable
     approximation at low redshift, at high redshift this time
     interval represents a significant fraction of the Hubble time and thus
     this assumption can induce large artificial offsets to the onset
     of star formation. Choosing higher density thresholds
     ($135\,h^2\,\rm cm^{-3}$) allows the entire cooling process to be
     followed, and the onset of star formation is then estimated to be
     at redshift $z\sim 12-16$. When isolated, rare, high-density
     peaks are considered, the chemical evolution is much faster and
     the first star formation episodes occur at $z\gtrsim 40$, almost 
     regardless of the choice of the density threshold.}
   {These results could have implications for the formation
     redshift of the first cosmological objects, as inferred from
     direct numerical simulations of mean-density environments and 
     studies of the reionization history of the universe.}


   \keywords{Cosmology: theory - early structure formation}


   \maketitle


\section{Introduction}\label{Sect:intro}
Understanding primordial structure formation is one of the fundamental
issues of modern astrophysics and cosmology. There is wide
agreement that not only consists the universe of ordinary ``baryonic''
matter but also a large fraction of unknown ``dark'' matter, whose effects
are only gravitational.
Baryonic matter appears to
constitute only a small fraction of the total cosmological matter
content with a present-day density parameter $\Omega_{0,b}=~0.0441$
compared to $\Omega_{0,m}=~0.258$ \cite[]{wmap5_2008}.
Since the universe is observed to have zero curvature, i.e. to have a
total density parameter $\Omega_{0,tot}=~1$, these
data imply that an additional density term exists $\Omega_{0,\Lambda}=~0.742$.
This is probably related to the so-called ``cosmological constant''
\cite[]{Einstein1917}, or, as initially suggested by
\cite{ratra1988}, \cite{Wetterich1988}, \cite{brax1999} and \cite{peebles2003},
to other kinds of unknown ``dark energies'', whose effects on early
structure formation history have been studied by e.g.
\cite{Maio_et_al_2006} with numerical simulations and by
\cite{Crociani_et_al_2008} with analytical calculations.\\
The existence of non-baryonic matter was suggested several
decades ago, and structure formation models based on the growth of
primordial gravitational instabilities \cite[]{Peebles1974,WhiteRees1978} were
developed following the early work by \cite{GunnGott1972}.\\
Hydrodynamical simulation codes 
[the first dating back to \cite{Evrard1988} and \cite{Hernquist_Katz_1989}]
have become a powerful tool, but because of computational limitations, 
plausible subgrid models have always been required to take into account star formation events
\cite[e.g. ][]{CenOstriker1992, Katz1992, Katz_et_al_1996, SpringelHernquist2003, DallaVecchia_Scahye_2008, Schaye_DallaVecchia_2008}.
These simulations model the converging gas infall into dark-matter
potential wells, by following the gas that becomes shock heated and
subsequently cools by atomic and/or molecular cooling. Given the many
orders of magnitude (in scale and density) spanned, it is computationally
extremely challenging to simulate the process down to the formation of 
single stars.\\
The gas physics in structure formation simulations has been typically
approached with either lagrangian smoothed particle hydro-dynamics (SPH) or Eulerian mesh codes.
A particular subclass is constituted by adaptive mesh refinement
(AMR) codes, which allow further decomposition of the mesh around
high-density regions, achieving a higher resolution.
The main advantage of the SPH approach is its ability to follow
self gravity in detail, while hydrodynamical instabilities are usually captured by mesh codes.\\
To account for star-formation episodes, both SPH and mesh
schemes rely on specific assumptions.
The prescriptions applied in mesh codes
\cite[e.g. ][]{CenOstriker1992, Inutsuka_Miyama_1992,Truelove_1997}
 usually assume that the star formation rate is proportional to the density of  overdense gas, while those used in SPH codes
\cite[e.g. ][]{Katz1992, Bate_Burkert_1997, SpringelHernquist2003}
are based on the existence of a density threshold above which the gas
is gradually converted into stars.
Here we make use of SPH simulations.\\
The typical timescales involved in the process of gas condensation
are the free-fall time, $t_{ff}$, and the cooling time, $t_{cool}$.
Gas condensation is expected to take place only if
$t_{cool}<t_{ff}$.\\
The free-fall time is defined as
\begin{equation}\label{t_ff}
t_{ff} = \sqrt{\frac{3\pi}{32 G\rho}},
\end{equation}
where $G$ is the universal gravitational constant and $\rho$ the density of
the medium; the numeric factor $(3\pi/32)^{1/2}$ is exact for
spherical symmetry only.
The cooling time is defined as
\begin{equation}\label{t_cool}
t_{cool} = \frac{3}{2} \frac{nk_B T}{\mathcal{L}(T,n_i)},
\end{equation}
where $n$ is the number density of the gas, $k_B$ the Boltzmann
constant, $ T$ the temperature, and $\mathcal{L}(T,n_i)$ the cooling
function (energy emitted per unit time and volume), which is dependent on both
temperature and number densities, $n_i$, of the species constituting
the gas.
In the low-density limit\footnote{
This widely-used approximation is appropriate as, according to the
classical spherical ``top-hat'' model, a virialized object has a total
mass density of $18\pi^2$ times the critical density, which
corresponds, on average, to a total number density of $\sim 2\,
h^2\,\rm cm^{-3}$
at $z\sim 15$, for a WMAP5 cosmology and a mean molecular
weight $\mu\simeq 1$.
The transition to a high-density statistical equilibrium
regime happens at critical number densities of $\sim\rm
10^4\,cm^{-3}$.
},
for two-body interactions, between particles $x$ and $y$,
$\mathcal{L}$ can be written as
\begin{equation}\label{lambda}
\mathcal{L}(T,n_x,n_y) = \Lambda (T) n_x n_y
\end{equation}
with the quantum-mechanical function $\Lambda (T)$ depending
on the temperature of the species considered, $n_x$ and $n_y$
\cite[see for example][]{Maio_et_al_2007}.
At $ T \geq 10^4$~K, the cooling is dominated by collisions of
hydrogen atoms, which is the most abundant species in nature
-- about $93\%$ in number fraction -- and $\mathcal{L}$ scales
approximatively as $n_H^2$ (we indicate with $n_H$ the hydrogen
number density).\\
The physical conditions in which the first structures form 
are characterized  by a primordial chemical composition: 
mostly hydrogen, deuterium, helium, and some simple
molecules, e.g. H$_2$ and HD.\\
The primordial sites in which the first stars form are thought to be
small dark-matter haloes with masses $\sim 10^6\,\rm M_{\odot}$ -- as
expected from predictions based on self-similar gravitational
condensation and chemical evolution
\cite[e.g. ][]{Tegmark_et_al_1997,TrentiStiavelli2009}
-- and virial temperatures $ T_{vir}\stackrel{<}{\sim} 10^4$~K.
Once they are born, they illuminate the universe and mark the end of
the ``dark ages''.
The radiation propagates in the vicinity of the individual sources and
the impact on the subsequent structure formation \cite[]{RGS2002a,
  RGS2002b, RGS_astroph2008} can be very 
significant leaving imprints by a means of feedback effects 
\cite[see][ for a review]{Ciardi_ferrara_2005}.\\
The low virialization temperatures of primordial
haloes  are enough neither to excite nor to ionize hydrogen and
the lack of any metals means that the gas can cool and eventually
form objects only {\it via} molecular transitions \cite[ for a detailed
study of the cooling efficiency in different
regimes]{SaZi67, Peebles_Dicke68, HM79, Maio_et_al_2007}.
They have rotational energy separations 
with excitation temperatures below $10^4$~K, and therefore it is
possible to collisionally populate their higher levels with the consequent
emission of radiation and resulting gas cooling. Since the
molecular energy-state separations are typically smaller than the
atomic ones, cooling will of course be slower, but still capable of bringing
the temperature down to $\stackrel{<}{\sim} 10^2$~K
\cite[]{Yoshida_et_al_2003, Omukai_Palla_2003, Yoshida_et_al_2006,
Gao_et_al_2007}.\\
To follow the entire process of structure and star formation
in numerical simulations, one should implement
the entire set of chemical reactions and hydrodynamical equations
and from those calculate the abundance evolution and the corresponding
cooling terms.
In practice, performing these computations is very expensive and time
consuming and it becomes extremely challenging to follow the formation
of structures from the initial gas infall into the dark-matter
potential wells to the final birth of stars.
Nevertheless, efforts are being made in this direction
\cite[e.g. ][]{ABN2002,BrommLarson2004,Yoshida2007,Whalen_et_al_2008}.\\
For this reason, more practical, even if sometimes
coarse, simple models are adopted.
In brief, star formation relies on semi-empirical and numerical
recipes based on chosen criteria to convert gas into stars and obtain
the star formation rate, carefully normalized to fit observational
data at the present day.
In particular, in SPH approaches a single particle represents a
population of stars with assigned mass distribution.
The standard method used is to assume that once the gas has
reached a given density threshold it automatically forms
stars\footnote{
To reduce the computation time for calculations of fragmentation sometimes, particles with densities above the threshold are replaced by `sink' particles \cite[][]{Bate_et_al_1995}.
}
[e.g \cite{CenOstriker1992}, \cite{Katz1992}, \cite{Katz_et_al_1996} and the popular \cite{SpringelHernquist2003} model, inspired by the previous works],
regardless of the time between the moment when the threshold is
reached and the effective run-away collapse, which typically takes
place at densities $\sim 10^2-10^{4}\,\rm cm^{-3}$.
While this is a reasonable approximation at low redshift, in ``average''
regions of the universe at
high redshift this time interval represents a significant fraction of the
Hubble time and thus the assumption can induce large artificial
offsets on the onset of star formation and influence the evolution in
the derived star formation rate.
Thus, extrapolations to high redshifts of the low-density thresholds 
(few $\rm 1^{-2}~cm^{-3}$) used to model the star formation rate in the 
low-redshift universe, may not always be justifiable.
For this reason, high-redshift applications require higher resolutions
and a higher density threshold.\\
In this paper, we are interested in modeling star formation as a global process 
in regions of mean density in the universe, not directly in the very first 
stars, which instead form in highly overdense, isolated regions.
In particular, we discuss the importance of the choice of the density
threshold for star formation in simulations of early structure formation.
We present a criterion to choose this threshold (Section 
\ref{sect:dth}) and some test cases based on 
high-resolution simulations (Section \ref{sect:sims}
and \ref{sect:results}). Then we present our results and conclusions
(Section \ref{Sect:conclusion}).


 \section{Threshold for star formation}\label{sect:dth}
According to the usual scenario of structure formation, the Jeans
mass \cite[]{Jeans1902} is the fundamental quantity that allows us to
distinguish collapsing from non-collapsing objects, under gravitational
instability.
For a perfect, isothermal gas, it is given by
\begin{equation}
M_J
=\frac{\pi}{6}\left( \frac{k_B T}{\mu m_H G} \right)^{3/2} \rho^{-1/2},
\end{equation}
where $m_H$ is the mass of the hydrogen atom, and $T$ and
$\rho$ are the temperature and density of the gas, respectively.
At very high redshift ($z\sim 30-20$), typical haloes have masses of
$\sim 10^5-10^6\, \rm M_\odot$,
which can increase up to $\sim 10^8-10^9\,\rm M_\odot$ by $z\sim 10$.\\
As mentioned in the introduction, the density threshold for star formation
in numerical simulations is typically fixed to some
constant value, irrespective of the simulation resolution.
However, it would be desirable to have a star formation
criterion that allows us to reach scales that fully resolve the Jeans
mass.\\
The SPH algorithm implicitly imposes a minimum mass resolution limit,
because to compute the different physical
quantities, a fixed number of neighbours (i.e. number of particles
within the smoothing length\footnote{
The definition of smoothing length may vary from authors to authors. 
It is sometimes meant to be the width of the SPH smoothing kernel, 
but other times the length scale on which the SPH smoothing kernel becomes zero.
}
, $h$) is used.
This also induces a minimum resolvable mass, which is the total mass of
neighbouring particles.
This is particularly important in SPH simulations of
cosmological structures and galaxy formation, because only
if the minimum resolvable mass is far smaller than the Jeans mass, it
is possible to ensure that the results are not affected by numerics
nor by the details of the implementation adopted
\cite[][]{Bate_Burkert_1997}.
Otherwise, unresolved, Jeans unstable clumps can easily be found to exhibit
unphysical behaviour (e.g. over-fragmentation problem in low-resolution simulations). Furthermore, it was shown
\cite[][]{Navarro_White_1993,Bate_Burkert_1997} that the minimum
number of particles needed to obtain reasonable and converging results
is about twice the number of neighbours ($\sim 10^2$ particles).\\
If $M_{res}$ is the gas mass resolution of a given
simulation, we can assume that:
\begin{equation}
M_J = N M_{res},
\end{equation}
where $N \gg 1$ and that the critical threshold is
\begin{eqnarray}\label{crit_th}
\rho_{th}& = &
\frac{\pi^2}{36 N^2 M_{res}^2} \left( \frac{k_B T}{\mu m_H G} \right)^3\\
&\simeq & \label{crit_th2}
\frac{1.31\cdot 10^{-13}}{N^2} \left(
\frac{M_{res}}{{\rm M}_{\odot}}\right)^{\!-2} \!\!\left(
\frac{T}{10^3~{\rm K}}\right)^{\!3} \!\!\left( \frac{1}{\mu}\right)^{\!3}\,\,\rm [g\,cm^{-3}].
\end{eqnarray}
For $M_{res}=10^2~{\rm M}_{\odot}$, $T=10^3~\rm K$, and $\mu=1$ and when using
$N=10^2$ gas particles, 
one has $\rho_{th}\sim 10^{-21}\,\rm g\,cm^{-3}$, corresponding to a
physical number density of $\sim 10^{2}\,\rm cm^{-3}$. 
The above equations should be considered as a guideline to estimating the
density threshold.
We note that we largely fulfill the \cite{Bate_Burkert_1997}
conditions, because the neighbour number in our simulations is $N_{neigh}=32$ {\bf(see Section \ref{sect:sims})}, so we resolve the Jeans mass with about three times the number of neighbours ($N\simeq 3N_{neigh}$).
As already mentioned however, commonly adopted density thresholds are
for practical reasons usually chosen to be of order of 
$\sim 10^{-1}\,\rm cm^{-3}$ or less
\cite[][]{Katz_et_al_1996,SpringelHernquist2003,Cecilia_et_al_2005,Governato_et_al_2007,TFS2007,Schaye_DallaVecchia_2008}.
For example, \cite{Wiersma_2009} used a
number density threshold of $10^{-1}\,\rm cm^{-3}$,
while the gas-particle mass was close to be the Jeans mass and the
gravitational softening of order of the Jeans length.\\
In general, in simulations of both cosmic structure and galaxy formation, it is quite hard to fully resolve the Jeans mass of the collapsing fragments, and star formation is often assumed to occur while the gas falling into the dark-matter potential wells is still heating up.
The \cite{Bate_Burkert_1997} requirement is not usually satisfied, since the main goal is usually not to follow the entire process of collapse and fragmentation, but to obtain a qualitatively representative sample of the cosmological evolution.\\
A high value for the threshold is also important to capture the relevant
phases of cooling.
In the following, we show that molecule radiative losses, at
temperatures of $\sim 10^3-10^4\,\rm K$ where they can balance the
heating of the infalling gas, produce an isothermal state and a
subsequent cooling regime.
For a region of mean density, the time spent by the gas in the 
isothermal state can be a substantial fraction of the Hubble time.
Therefore, it is important for the threshold to be
on the right-hand side of the peak in the phase-diagram (i.e. at densities
higher than the isothermal regime), so that the delay between reaching the
threshold and the true star formation is negligible (see Section 4).\\
In the following, 
we investigate the effect of different choices of
star formation thresholds at high redshift, 
describe the simulations performed, 
and discuss the results obtained.

\begin{table*}
\centering
\caption[Simulation parameters]{
Parameters adopted for the simulations. 
}
\begin{tabular}{lcccccccccr}
\hline
\hline
Model & Number of& $M_{gas}\rm$  & $M_{dm}\rm$ &
$\Omega_{0M}$ & $\Omega_{0\Lambda}$ & $\Omega_{0b}$ & $h$ & $\sigma_8$
& n & SF threshold \\
 &gas+dm particles & $[\rm M_{\odot}/{\it h}]$ & $[\rm M_{\odot}/{\it h}]$ & & & & & & & $[h^2\rm cm^{-3}]$\\
\hline\\
wmap5-ht   &$2\times 32768000$ & 128 & 621 & 0.258 & 0.742 & 0.0441& 0.72  & 0.8 & 0.96 &135.0\phantom{x}\\
wmap5-lt   &$2\times 32768000$ & 128 & 621 & 0.258 & 0.742 & 0.0441& 0.72  & 0.8 & 0.96 &0.2\phantom{x}\\
std-ht 	   &$2\times 32768000$ & 116 & 755 & 0.300 & 0.700 & 0.0400& 0.70  & 0.9 & 1.00 &135.0\phantom{x}\\
std-lt     &$2\times 32768000$ & 116 & 755 & 0.300 & 0.700 & 0.0400& 0.70  & 0.9 & 1.00 &0.2\phantom{x}\\
zoom-std-ht&$2\times 41226712$ & 3.9 & 25.6 & 0.300 & 0.700 & 0.0400& 0.70  & 0.9 & 1.00 &135.0\phantom{x}\\
\hline
\hline
\label{tab:sims}
\end{tabular}\\
\flushleft
The columns (from left to
right) specify: name of the run,
number of particles used,
gas-particle mass,
dark-matter-particle mass,
$\Omega_{0,m}$,
$\Omega_{0,\Lambda}$,
$\Omega_{0,b}$,
$h$,
$\sigma_8$,
spectral index,
star formation density threshold.
\end{table*}
\section{Simulation set-up}\label{sect:sims}
To study the effect of different threshold prescriptions
on the onset of star formation, we completed very high
resolution, three-dimensional, hydrodynamic simulations including
non-equilibrium atomic and molecular chemistry, star formation, and
wind feedback.\\
We used the code Gadget-2 \cite[]{Springel2005}
in its modified form, which includes stellar evolution and metal
pollution \cite[]{TBDM2007}, primordial molecular chemistry (following
the evolution of  e$^-$, H, H$^+$, He, He$^+$, He$^{++}$, H$_2$,
H$^+_2$, H$^-$, D, D$^+$, HD, $\rm HeH^+$), and fine structure metal
transition cooling (O, C$^+$, Si$^+$, Fe$^+$) at temperatures lower
than $10^4$~K \cite[]{Maio_et_al_2007,Maio_et_al_2008}.
We perform hydro-calculations by fixing the number of SPH neighbours to
$N_{neigh}=32$.\\
The simulations have a comoving box size of $L=1~\rm Mpc$
and sample the cosmological medium with a uniform realization of
$320^3$ particles for both gas and dark-matter species (for a total
number of $2\times 320^3$ particles).
The resulting gas-particle mass is of the order of $\rm 10^2\,M_\odot$,
which is consistent with the discussion in the previous section.\\
We note that this configuration enables us to easily resolve the Jeans
length for shock heated/cooling cosmic gas (the Jeans length for
gas with $\rm T\sim 10^{4}\,K$ and $\rm \rho\sim
10^{-25}-10^{-24}\,g/cm^3$ is $\sim\rm 3 - 1\,kpc$,
much longer than the comoving gravitational softening\footnote{
The comoving gravitational softening is usually estimated as $1/20$ or
$1/30$ the mean inter-particle separation.
}
of $\rm\sim 0.1\,kpc$).\\
The initial conditions (set at redshift $z=100$) are
generated with a fast Fourier transform  grid of $N_{mesh} =~320$
meshes and a maximum wave-number (Nyquist frequency)
\begin{equation}
k_{Nyquist}=
\frac{2\pi N_{mesh}}{2L} \simeq\rm 1~kpc^{-1}\nonumber
\end{equation}
(i.e. a minimum wavelength of $2L/N_{mesh}\simeq\rm~6.25~kpc$),
so that, for each wave-number, $\|{\bf k} \|<k_{Nyquist}$.\\
We will refer to this sampling as ``mean region''.\\
We considered two different sets of cosmological parameters:

\begin{itemize}

\item
{\it standard model}: $\Omega_{0,m}=0.3$, $\Omega_{0,\Lambda}=0.7$,
$\Omega_{0,b}=0.04$, $h=0.7$, $\sigma_8=0.9$ and $n=1$, where the
symbols have the usual meanings.
The corresponding dark-matter and gas-particle masses are $\sim
755\,{\rm M}_\odot/h$ and $\sim 116\,{\rm M}_\odot/h$, respectively.

\item
{\it WMAP5 model}: data from 5-year WMAP (WMAP5) satellite
\cite[]{wmap5_2008} suggest that
$\Omega_{0,m}=0.258$,
$\Omega_{0,\Lambda}=0.742$,
$\Omega_{0,b}=0.0441$,
$h=0.72$,
$\sigma_8=0.796$, and
$n=0.96$.
In this case, the corresponding
dark-matter and gas-particle masses are
$\sim 621\,{\rm M}_\odot/h$ and $\sim 128\,{\rm M}_\odot/h$, respectively.

\end{itemize}

\noindent
Following the discussion in the previous sections, we also consider
 two different values for the star formation density threshold:

\begin{itemize}

\item
 a {\it low-density} threshold of $0.2\, h^2\rm cm^{-3}$
 (physical), compatible with the one adopted in the Gadget code and the
 ones widely  used in the literature \cite[for example][]{Katz_et_al_1996,SpringelHernquist2003,TFS2007, Pawlik_et_al_2009};

\item
 a {\it high-density} threshold of $135\,h^2\rm cm^{-3}$ (physical),
 as computed from Eqs. (\ref{crit_th}) and (\ref{crit_th2}). This
 value is adequate for modelling atomic processes even in small $\sim
 10^5\,\rm M_\odot$ haloes at $z\sim 20$.
 Moreover, this threshold typically falls in density regimes where
 cooling dominates over heating, allowing us to properly resolve
 gas condensation down to the bottom of the cooling branch.
\end{itemize}

\noindent
A summary of all the simulation features is given in Table \ref{tab:sims}.
We denote with the labels ``std'' and
``wmap5'' the runs with standard and WMAP5 cosmology, respectively, and
with ``lt'' and ``ht'' the runs with low- and high-density thresholds,
respectively.\\
We note that the \citet{SpringelHernquist2003} model used here to
describe the star formation process is strictly applicable only as long
as more than one star per SPH particle is present, i.e. each SPH
particle is considered as a 'simple stellar population' with a given
mass distribution. Although some studies
\cite[][]{Yoshida_et_al_2003,BrommLarson2002,BrommLarson2004}
 seem to indicate (or assume) that the very first episode of 
star formation could result in a single, very massive star per halo,
this should apply preferentially to very high redshift, high density,
isolated objects. In any case, the exact shape of the IMF of primordial
stars \cite[e.g. ][]{SS1953,Larson1998,NakamuraUmemura2001,Omukai_Palla_2003}
is still a matter of speculation and lively debate.
For this reason and because we are interested mainly in the global star
formation process, we used the \citet{SpringelHernquist2003} model 
[as \cite{TFS2007} also did to describe both a primordial top-heavy and a more standard star formation mode]
and allowed the IMF to be a
free parameter (although in the test cases reported here we always
adopt a top-heavy IMF). The aim of this paper is to investigate the
effects of the density threshold on star formation, and we leave
discussion on the IMF to future work.\\
Finally, to investigate primordial star formation events in
local high-density regions, we perform a very high-resolution numerical
simulation of a rare high-sigma peak with comoving radius $\sim
140\,{\rm kpc}/h$.
This region is selected using the zoomed initial condition technique
on a $\sim 10^9\,\rm M_\odot$ halo formed in a dark-matter-only
simulation \cite[]{Gao_et_al_2007}\footnote{
We use the ``R4'' initial conditions presented there.
}.
We divided each particle into gas and dark-matter component, according to
the standard model parameters.
The resulting gas-particle mass is $\sim 4\,{\rm M}_{\odot}/h$ and dark
matter particles have a mass of $\sim 26\, {\rm M}_\odot/h$
(in Table \ref{tab:sims}, this simulation is labelled ``zoom-std-ht'').\\
By a quick comparison of the different parameters adopted, we expect that, once the density threshold has been fixed, the standard cosmological mean-region simulations will show earlier structure formation episodes with respect to the corresponding wmap5 ones. This is because they have higher spectral parameters and higher matter content. The high-density region is a biased over-dense region already at early times, and therefore, its evolution is expected to be much faster.


\section{Results}\label{sect:results}
We present the results of the simulations with the sets of
parameters described above.
We discuss first the mean region of the universe (Section
\ref{sect:cosmo}) and then the high-density region (Section
\ref{sect:zoom}).

\begin{figure*}
\centering
%
%
\includegraphics[width=\textwidth]{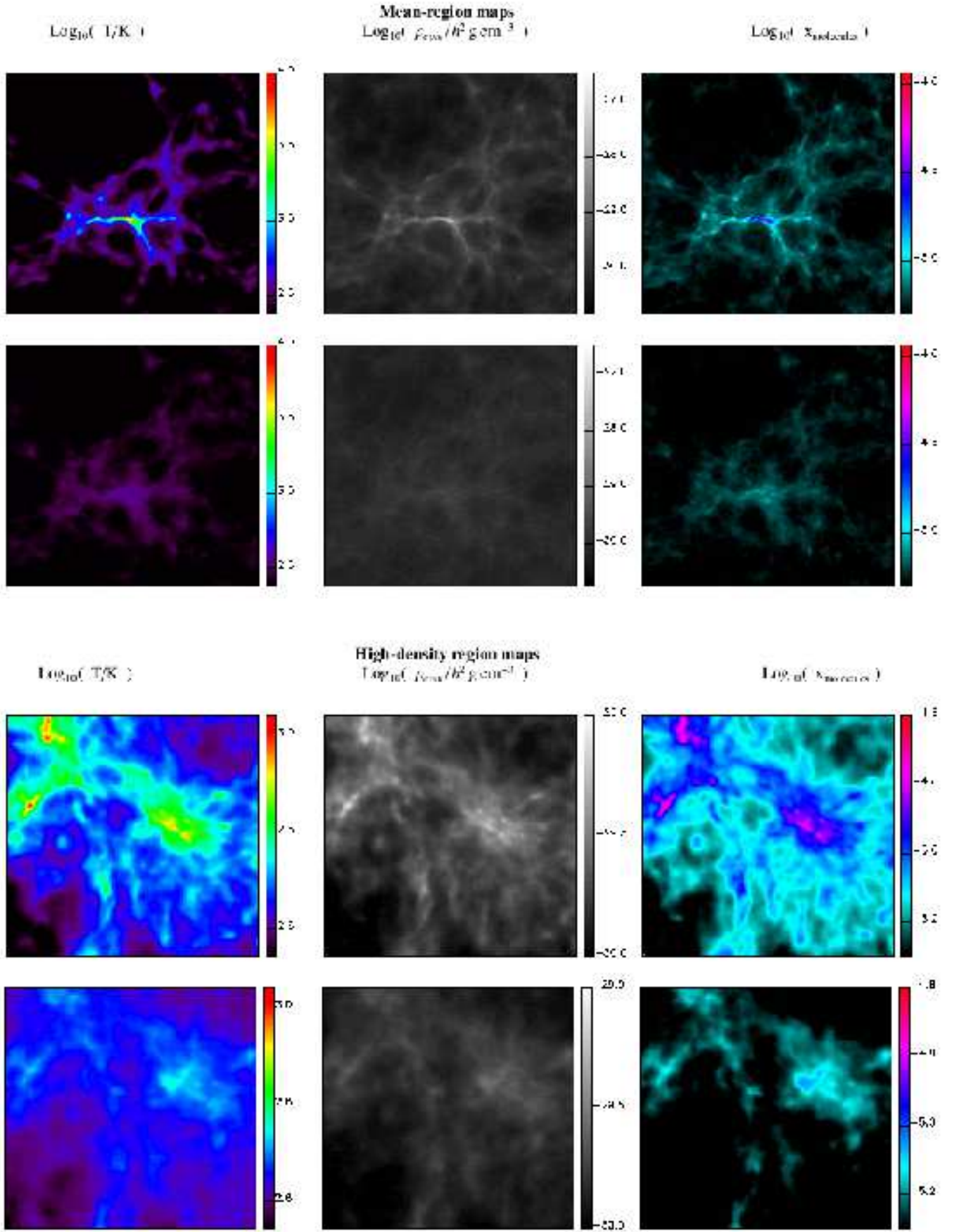}
\caption[Cosmological maps]{\small
First, second, and third column are respectively 
temperature, density, and molecule maps.
The first two rows refer to the mean-region simulation at redshift
12.17 (top) and  30.16 (bottom).
The box size is 1 Mpc comoving.
The last two rows refer to  the high-density region at redshift
50 (top) and 70 (bottom).
The region size is $\sim 140\,\rm kpc/{\it h}$ comoving.
All quantities are smoothed on a $276-$pixel side grid.
}
\label{fig:maps}
\label{fig:maps_zoom}
\end{figure*}

\subsection{Mean-region simulation}\label{sect:cosmo}
\begin{figure}
\centering
\includegraphics[width=0.5\textwidth]{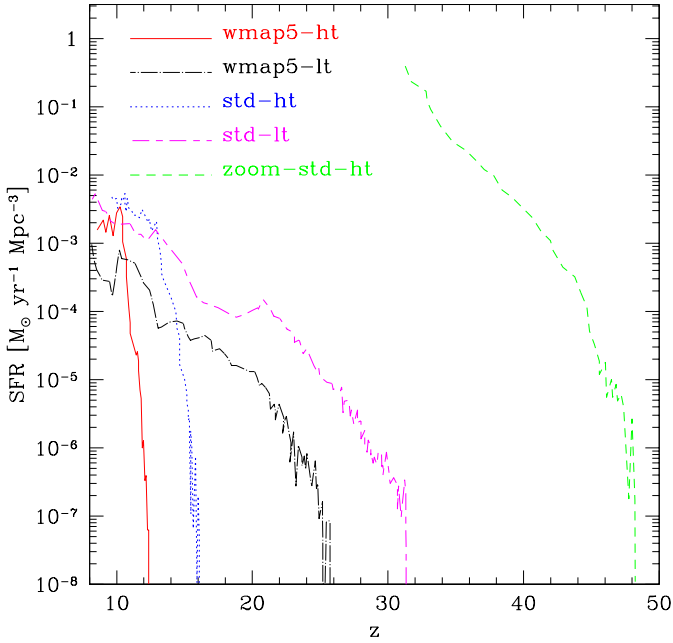}
\caption[SFR]{\small
Star formation rate as a function of redshift for the different
models, from left to right: 
WMAP5 cosmology and high-density threshold (solid red line),
standard cosmology and high-density threshold (dotted blue line),
WMAP5 cosmology and low-density threshold (dot-dashed black line), 
standard cosmology and low-density threshold (long-dashed-short-dashed
magenta line).
The green short-dashed line refers to the simulation of the high-density
region with standard parameters and high-density threshold.
}
\label{fig:sfr}
\end{figure}

Our reference run is the wmap5-ht model with initial composition given
by the values quoted in \cite{GP98}\footnote{
We assume a primordial neutral gas with residual electron and H$^+$
fractions
$x_{\rm e^-}\simeq x_{\rm H^{+}}\simeq 4\cdot 10^{-4}$,
H$_2$ fraction $x_{\rm H_2}=10^{-6}$,
H$_2^+$ fraction $x_{\rm H_2^+}=3\cdot 10^{-21}$,
D fraction $x_{\rm D}=3.5\cdot 10^{-5}$,
HD fraction $x_{\rm HD}= 7\cdot 10^{-10}$,
D$^+$ fraction $x_{\rm D^+}=4\cdot 10^{-9}$,
HeH$^+$ fraction $x_{\rm HeH^+}=   10^{-14}$.
} at $z=100$.
We show some evolutionary stages in Fig. \ref{fig:maps} (upper set
of panels).
In the maps, the first column refers to temperature, the second to gas
density and the third to molecular fraction at
$z= 30.16$ and $z=12.17$, respectively.
The creation of new molecules is clearly evident, together with the
related growth of structures. More specifically, as time passes, one
can see the heating undergone by the gas in dense regions, because of
structure formation shocks. The temperature increases from a few 
hundreds Kelvin in the low-density regions, to $\sim 10^4\,\rm K$ in
the denser regions.
In the meantime, the molecular fraction also evolves accordingly up
to values higher than $10^{-4}$. Soon after, the production of
molecules increases rapidly (up to $\sim 10^{-2}$) aiding the star
formation process, which, for this simulation, starts at $z\simeq 12$.
\begin{figure}
\centering
\includegraphics[width=0.5\textwidth]{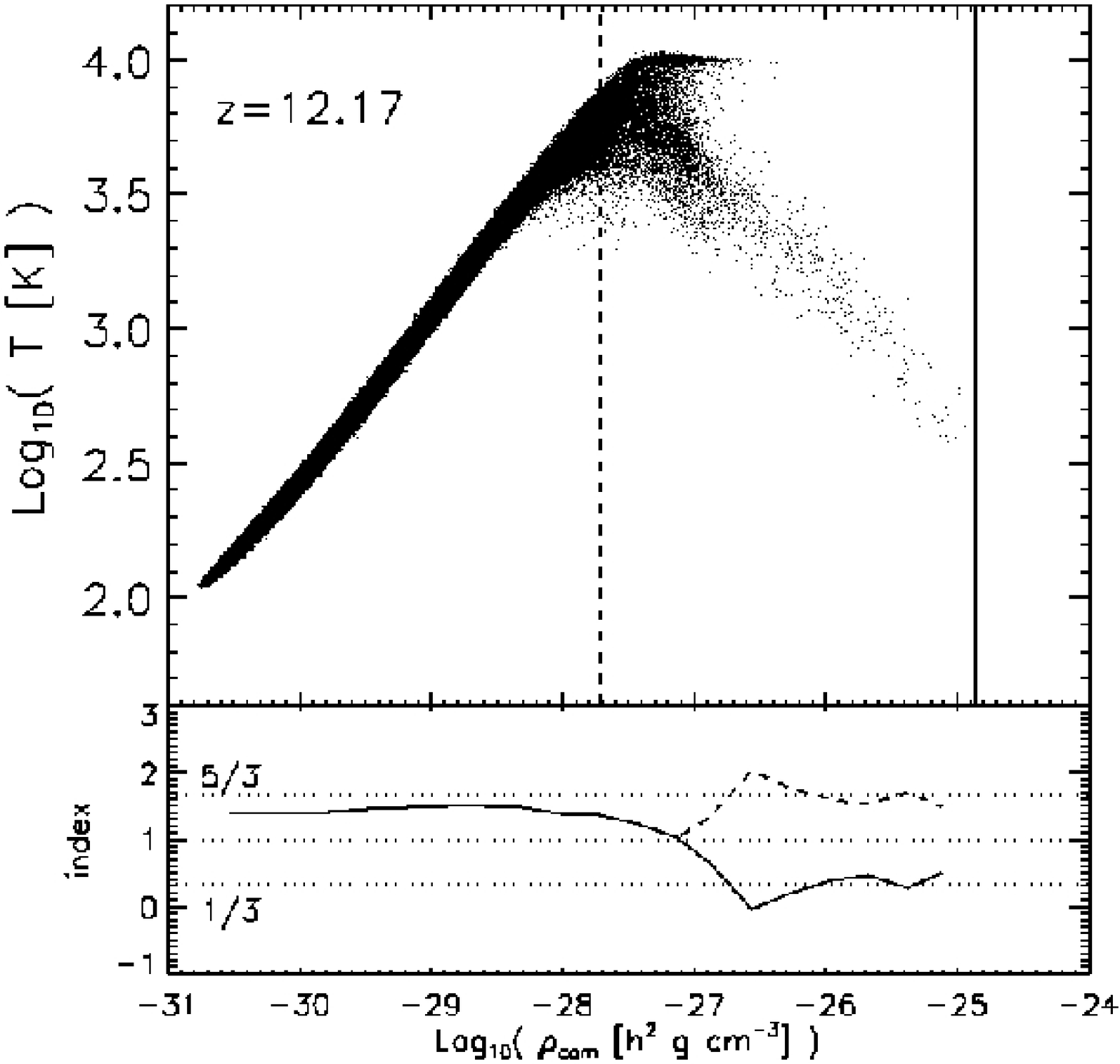}
\caption[Phase diagram]{\small
{\it Upper panel}:
phase diagram at redshift $z=12.17$ (just before the onset of
star formation) for the wmap5-ht simulation.
The vertical straight lines indicate a
low physical critical density threshold of $0.2~h^2\rm cm^{-3}$
(dashed line) and a higher physical critical density threshold of
$135\,h^2\rm cm^{-3}$ (solid line).
{\it Lower panel}:
average effective index computed over the whole range of
densities. The three horizontal dotted lines show
values of 5/3, 1 and 1/3, respectively from top to bottom. 
The solid line shows $\alpha$ and the dashed line shows $\gamma$ (see
text for definitions).
}
\label{fig:phase}
\end{figure}
\begin{figure}
\centering
\includegraphics[width=0.52\textwidth]{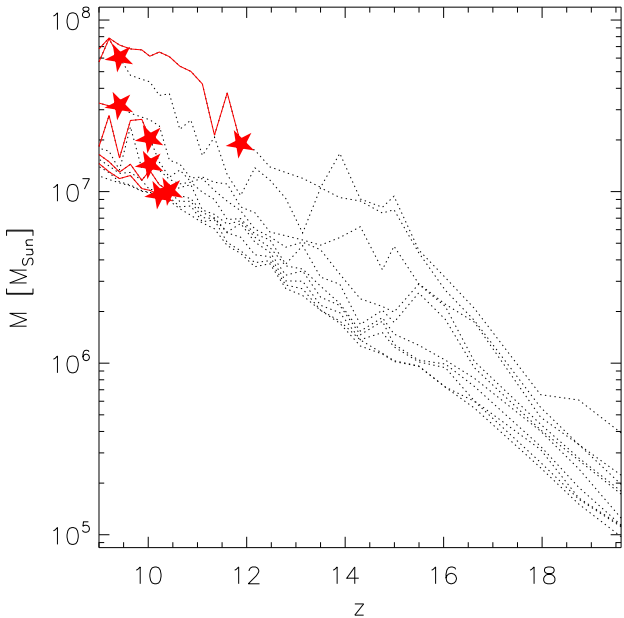}
\caption[Evolution of primordial haloes]{\small
Evolution of the ten most massive haloes in the wmap5-ht cosmological 
simulation (dotted lines).
The redshift at which the first star forms in each halo is indicated by the
filled star symbols. After that, star formation continues along the
solid lines.
}
\label{fig:haloes}
\end{figure}\\
In Fig. \ref{fig:phase}, we show the phase diagram (comoving
density versus temperature)
at redshift $z\simeq 12$, i.e. just before the onset of star formation.
The low-density gas, which is shock-heated by the collapse of the first
primordial haloes, is seen on the left side of the panel.
Starting from values for a temperature of $\sim 10^2$~K, the gas is
progressively heated to $\sim 10^4$~K and moves along the rising
branch.
At this stage, collisions become more frequent due to the higher
temperature.
The upper energy levels of particles become excited and the subsequent
de-excitation is accompanied by the emission of radiation. This
effect is negligible at low densities, because collisions are rare
and the fraction of energy converted into radiation is small. When the
density increases, the cooling becomes comparable to the heating and
an isothermal regime with no significant net change in the temperature
is reached. This appears at the tip of the phase diagram (and
in the behaviour of the effective index, as discussed below), at
$T\sim 10^4\,\rm K$, where the cooling is dominated by atomic
Ly$\alpha$ transitions and accompanied by runaway collapse.
At higher densities, radiative losses overtake heating and induce a 
fast cooling phase (dominated by molecules, mostly H$_2$).\\
The solid vertical line corresponds to the physical high-density star
formation threshold ($135\,h^2\,\rm cm^{-3}$) and, for comparison, we
also plot the dashed line for a physical number density of
$0.2\,h^2\,\rm cm^{-3}$.
We stress that by adopting a low-density threshold for star
formation one completely misses the isothermal and cooling part of
the phase diagram, and thus a correct modeling of the cooling
regions within the simulations. This can affect the onset of star
formation, particularly at high redshift, when the time needed for the gas 
to evolve from the low-density threshold to the high-density threshold
($\sim 2\cdot~10^8\,\rm yr$) can be a substantial fraction of
the Hubble time ($\sim 4\times 10^8\,\rm yr$ at $z\sim 12$). We note that
the time elapsed between the attainment of the isothermal peak in the
phase diagram and the end of the cooling branch is $\sim
6\times~10^7\,\rm yr$.
The evolution that follows the end of the cooling branch
is characterized by the formation of a dense core, which accretes gas
on free-fall timescales \cite[]{Yoshida_et_al_2006}.
This phase has a very short duration ($\sim 10^6-10^7\,\rm yr$) during which the
central densities increase to $\sim 10^{16}\,\rm cm^{-3}$.
The problem is less severe at lower redshift, when the Hubble time
becomes of the order of several Gyr.\\
The density and temperature behaviour can also be described
by an effective index\footnote{
By effective index of the gas, $\gamma$, we mean $P\propto
\rho^\gamma$, with $P$ pressure and $\rho$ density.
This is simply related to the politropic index.
},
which depends on the physical conditions of the gas regime
considered.
In the lower panel of Fig. \ref{fig:phase}, we plot the effective
index as a function of density.
The solid line refers to the value
$\alpha \equiv 1 +~(d T/T)/ (d \rho / \rho)$,
which takes into account changes in the sign of the temperature
derivative, distinguishing the heating regime ($\alpha>~1$) from the
cooling regime ($\alpha<~1$) .
The dashed line refers to
$\gamma \equiv 1 +~|(d T/T)/(d \rho / \rho)|$, 
so that $\gamma$ is always $\geq 1$.
Dotted horizontal straight lines INDICATE values of $5/3$, $1$, and
$1/3$.
In correspondence with the isothermal peak in the $T-\rho$ plane,
it is $\alpha=\gamma=1$, which marks the transition from the heating
to the cooling regime.
At this stage, we expect the gas runaway collapse to begin and last
for the following cooling regime, at which point $\alpha$
oscillates around the value of 1/3.\\
In Fig. \ref{fig:haloes}, we plot the evolution of the ten most
massive haloes found in the simulation. We also show the redshift at which
stars are produced (filled star symbols) in each object.
The haloes are found using a friend-of-friend
algorithm with a linking length equal to $20\%$ of the
mean inter-particle separation.
Typical halo masses at redshift $z\sim 12$, when star
formation starts, are of the order of $10^7\,\rm M_{\odot}$
\cite[see also][]{Wise_Abel_2007,Wise_Abel_2008} and reach densities of $\sim
10^2\,\rm cm^{-3}$.\\
For comparison, we completed the same simulation using standard
cosmological parameters (std-ht run).
In this case, the overall picture is similar, but we detected 
a faster evolution, with earlier structure formation,
as expected from the previous discussion in Section \ref{sect:sims}.
The first star formation events are detected at redshift $z\sim 16$ in haloes  with masses $\sim 10^7\,\rm M_\odot$.\\
This can clearly be seen in Fig. \ref{fig:sfr}, where we plot the
star formation rate as a function of redshift for the different simulations
(to compute the star formation rate, we adopt the implementation
described by \cite{SpringelHernquist2003}).\\
The onset of star formation in the wmap5-ht
model (red solid line) is delayed compared to the std-ht model
(blue dotted line).
For the wmap5-lt (black dashed line) and std-lt (magenta
short-long-dashed line) models, star formation starts at $z\sim 25$
and 31, respectively.
Thus, at these high redshifts, even small changes in the cosmology can
be significant for the onset of star formation.
This is easily understood in terms of spectral parameters: the
standard cosmology has higher spectral index and normalization;
therefore, assigning more power on all scales with respect to WMAP5
values, leads to structure formation occurring much earlier.\\
The choice of the density threshold makes an
even larger difference to the onset of star formation.
In Fig. \ref{fig:sfr}, the rates corresponding to the
wmap5-lt (black dot-dashed line) and wmap5-ht (red solid line) show
that star formation starts at $z\sim 25$ and 12, respectively.
The major difference between low- and high-density
threshold models is that, in the former, the gas
reaches the critical density much earlier.
So, the redshift difference in the onset corresponds to the
time that the gas needs to move from the low- to the high-density
threshold (see Fig.~\ref{fig:phase}).\\
In addition, the simulations adopting the high-density
thresholds slightly overtake the respective low-threshold cases.
This happens because the former did not remove the gas at higher
redshifts, it accumulated and ended in delayed bursts of star
formation.
Later, the star formation rates were restored to the same level.\\
As already mentioned, the low-density threshold model is very commonly
used both in numerical and semi-analytical works, because it
does not require incorporation of molecular chemistry (the threshold
being lower than the typical densities at which molecules become 
efficient coolants) and therefore it is easier to implement and allows
faster simulations.
However, it can compromise the entire picture if the results are
extrapolated to high redshift, when molecules are the main coolants
and the time delay between the attainment of the low-density threshold
and the bottom of the cooling branch occupies a significant fraction of 
the Hubble time.

\subsection{High-density region simulation}\label{sect:zoom}
We show results for the high-density
region described in Section \ref{sect:sims} and initialized at
redshift $z=399$.\\
In this case, the physical number densities at the beginning of the
simulation are in the range $\sim 0.5-50\,h^2\,\rm cm^{-3}$ (at $z\sim
200$), with an average of $\sim 4 \,h^2\,\rm cm^{-3}$, higher than the
typical value adopted for the low-density threshold for star
formation.
Therefore, the conventional low-density model would
produce unreasonable star formation at $z\sim 200$.
To avoid this, it is common to add a further, additional,
{\it ad hoc} constraint, which allows star formation only if the
simulation over-densities are higher than a given minimum value~--
usually between $\sim 50$ and $\sim 100$ \cite[][ in Section 4.2, for example, suggest 55.7]{Katz_et_al_1996}.
Thus, in this case it is this additional constraint that determines
when the onset of star formation occurs, rather than the low-density
threshold.\\
We therefore ran a simulation with only a high-density threshold.
For the sake of comparison, we still used the value of $135\, h^2\,\rm
cm^{-3}$, although rigorously, following Eqs. (\ref{crit_th}) and
(\ref{crit_th2}), one should adopt a value $\sim 9\times 10^4\,h^2\,\rm
cm^{-3}$ for a 3.9 M$_\odot/h$ gas-particle mass.
Nonetheless, we checked that this choice does not affect our
conclusions, since the threshold is already  beyond the isothermal peak, in the
fast cooling regime, where the timescales are extremely
short ($\sim 10^6\,\rm yr$).
All the initial abundances are set according to the values suggested
by \cite{GP98}\footnote{
The initial abundances (at $z=399$) are consistent with a primordial
neutral plasma having residual electron and H$^+$ fractions of
$x_{\rm e^-}\simeq x_{\rm H^{+}}= 10^{-3}$,
H$_2$ fraction $x_{\rm H_2}=10^{-10}$,
H$_2^+$ fraction $x_{\rm H_2^+}=3\times 10^{-15}$,
D fraction $x_{\rm D}=3\times 10^{-5}$,
D$^+$ fraction $x_{\rm D^+}=3\times 10^{-8}$,
HD fraction $x_{\rm HD}=10^{-14}$,
HeH$^+$ fraction $x_{\rm HeH^+}=5.6\times 10^{-18}$}.\\
The simulation maps are shown in Fig. \ref{fig:maps_zoom}
(lower panels). As for the mean-density regions, they refer to
temperature, density, and molecular fraction at redshifts z=50 and 70.
As expected, we highlight that structure formation occurs far
earlier than the mean-density case.
Molecular abundances of $\sim 10^{-5}$ are reached faster than
for the mean-density region, where such a high fraction
is found only at $z\lesssim 30$.
Similarly, values of $\sim 10^{-4}$ are already reached at $z\sim
50-40$, rather than $z\sim 20$ -- see also discussion in Section
\ref{Sect:conclusion} and Eq. (\ref{t_cool_H2}).\\
\begin{figure}
\centering
\includegraphics[width=0.5\textwidth]{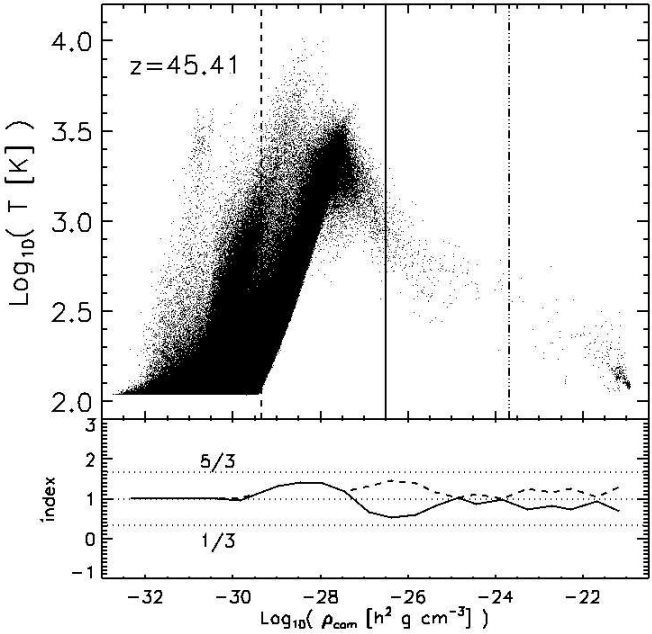}
\caption[Phase diagram]{\small
{\it Upper panel}:
phase diagram at redshift $z=45.41$ for the high-density region
simulation for a purely non-equilibrium chemistry run (i.e. without
star formation).
The vertical straight lines show physical critical-density threshold
of $0.2~h^2\rm cm^{-3}$ (dashed line), 
$135\,h^2\rm cm^{-3}$ (solid line) and 
$8.9\times 10^4\,h^2\rm cm^{-3}$ (dot-dashed line).
{\it Lower panel}:
average effective index computed over the whole range of densities.
The three horizontal dotted lines show values of 5/3, 1, and 1/3,
respectively, from top to bottom.
The solid line refers to $\alpha$ and the dashed line to $\gamma$ (see
text for the definitions).
}
\label{fig:phase_zoom}
\end{figure}
In Fig. \ref{fig:phase_zoom}, we show the phase diagram and
the behaviour of the effective index as a function of the comoving gas
density at redshift $z\simeq 45$.
Physical critical density thresholds of
$0.2~h^2\rm cm^{-3}$ (dashed line),
$135\,h^2\rm cm^{-3}$ (solid line), and
$\sim 9\times 10^4\,h^2\rm cm^{-3}$ (dot-dashed line)
are marked in the figure.
While the first two are the same as for the mean-density
region simulation, the last one corresponds to the value obtained using
Eqs. (\ref{crit_th}) and (\ref{crit_th2}).
To emphasize the different characteristics of the phase diagram compared
to the one obtained for the mean-density region, 
the plot was extended to densities higher than before.
The isothermal peak is reached at redshift $z \sim 50$.
Unlike the mean-density simulation, the gas does not spend 
time on the isothermal plateau, but cools very rapidly 
(in less than  $7\times 10^6\,\rm yr$) from $\sim
10^{3.5}\,\rm K$ to $\sim 10^2\,\rm K$ and condenses into comoving
densities of $\rho_{com}\sim 10^{-21}\, h^2\, \rm g/cm^3$.
The rapidity of these events is reflected in the lack of particles in
the intermediate stages of the cooling branch.\\
As before, we also plot the effective gas index.
The usual initial shock-heating behaviour and the following
cooling is recovered up to much higher densities. At the
bottom of the cooling branch, we find values of $\alpha$ that
oscillate around 1/3 and 1.
Since the last stages are quite fast, the low number of
particles present introduces some statistical noise, which is evident
in the plot.\\
With our choice of the threshold, star formation sets in at 
$z\sim 48$ (see Fig. \ref{fig:sfr}). The additional time
needed to reach the highest densities
at the bottom of the cooling branch is extremely short 
($\sim 10^6\,\rm yr$), so our choice ensures that the onset of star
formation is correctly estimated.
As there is no obvious, standard way of quantifying the star
formation rate in these simulations, we do this by
dividing the stellar mass formed at each
time-step by the volume of the gas contained in the high-density
region (a sphere of about $140~{\rm kpc}/h$ radius).\\


\section{Discussion and conclusions}\label{Sect:conclusion}
We have studied the effect of different choices of the density
threshold on the onset of cosmic star formation in numerical SPH
simulations
\cite[see][ for technical details]{Maio_et_al_2007,TBDM2007}.\\
In the literature, several studies are presented that follow the birth
of primordial stars in early protogalaxies
\cite[examples are][]{Yoshida_et_al_2006,Wise_Abel_2007}.
These are mainly focused on the initial phases of star formation and
the effects on the immediate surroundings, which typically do not
address more general issues such as the global star formation process
(i.e. in a region at mean density rather than in high density peaks),
metal enrichment, and IGM reionization \cite[e.g. ][]{BoltonHaehnelt2007}.
Here instead, we are interested in simulating the more global star formation
process in the high-redshift universe, capturing the relevant
timescales and physical processes, i.e. the atomic and molecular
physics that regulates the formation of primordial stellar population.\\
We have thus run simulations using initial conditions appropriate to 
a region of the universe with mean density and as a reference, using the zoom
technique, a high-density peak.\\
A basic process that leads to star formation, i.e. gas shock
heating up to
$\sim 10^3-10^4\,\rm K$ by infall into dark-matter haloes followed
by radiative losses due mainly to molecular collisional excitations,
is common to both scenarios.
The main difference is associated with the global dynamics and
timescales of the process.
In the rare high-sigma peak, 
because of the higher densities, chemical reactions are faster
and much more efficient with respect to the simulations of
mean-density initial conditions. Therefore, the molecular fraction
increases more rapidly, reaching a number fraction of $\sim 10^{-4}$ by
$z\sim 50-40$ (compared to $z\sim 20$, for the corresponding
mean-density case).
These values are sufficient to make collisional cooling to dominate over
heating and induce star formation episodes.\\
Density and temperature behaviour can be described by
an effective index that depends on the physical conditions
of the gas considered.
Roughly, it is isothermal during the transition from
the heating to the cooling regime, then it collapses
(see the effective index computed in the lower panel of Fig.
\ref{fig:phase})
until the bottom of the cooling branch is reached.
More quantitatively, when cooling is dominated by H$_2$,
the cooling time in Eq. (\ref{t_cool}) can be approximated as
\begin{equation}\label{t_cool_H2}
t_{cool}\simeq \frac{3}{2}\frac{k_BT}{\Lambda_{\rm H_2}(T)\,\,x_{\rm
    H_2}n_{\rm H}}
\end{equation}
where $x_{\rm H_2}$ is the H$_2$ number fraction, $\Lambda_{\rm
  H_2}(T)$ is the H$_2$ cooling function at temperature $T$, and the
other symbols have their usual meanings.\\
For gas at the beginning of the cooling branch, $T\sim 10^{3.5}\,\rm K$ and
$x_{\rm H_2}\sim 10^{-4}$, giving $~t_{cool}\sim 7\times~10^6 \,
n_{\rm H}^{-1}~$~yr ($n_{\rm H}$ in cm$^{-3}$).
In the mean-density case (see phase diagram in Fig.~\ref{fig:phase}), 
$n_{\rm H}\simeq 0.3\rm\, cm^{-3}$, while in the high-density region
(see phase diagram in Fig.~\ref{fig:phase_zoom}),
$n_{\rm H}\simeq 6\rm\,cm^{-3}$.
This translates into a characteristic cooling time of $\sim 2\times
10^7\,\rm yr$ for the former case and $\sim 10^6\,\rm yr$ for the
latter.\\
These estimates show the relevance of following the
full cooling branch when simulating star formation at high redshift in
regions of mean density, because the characteristic cooling times
are a substantial fraction of the Hubble time. This problem
is less severe for simulations of high-density peaks, in which the
timescales are much shorter.\\
If one considers a $10^5\,\rm M_\odot$ halo at $z\sim 40$, its
average gas number density is of the order of $\sim 20/\mu\,\rm
cm^{-3}$ (being $\mu$ the mean molecular weight), its virial temperature $\rm T_{vir}\sim 7\times 10^2\mu\,K$ and its cooling time $t_{cool}\sim  3\times 10^5- 3\times 10^8\,\rm yr$ for an H$_2$ fraction of $10^{-2}-10^{-5}$, respectively.
The same halo at redshift $z\sim 15$ would have an average gas number
density of $\sim 1/\mu\,\rm cm^{-3}$, $\rm T_{vir}\sim 5\times 10^2\mu\,\rm K$, and $t_{cool}\sim 7\times 10^6-7\times10^9\,\rm yr$.
This means not only that objects of similar mass cool (and thus harbor
star formation) on very different timescales according to their
environment, but also that gas in high-redshift haloes can collapse and fragment much faster than in low-redshift ones, because its cooling capabilities are more efficient.\\
For these reasons, though low-density thresholds can
be useful tools to reproduce empirical surface-density relations, such as
the Kennicutt-Schmidt law \cite[]{Kennicutt1998,SpringelHernquist2003,Schaye_DallaVecchia_2008},
physical insights can rely only on high-density thresholds.
Indeed, imposing star formation events before the isothermal peak
is reached could result in an artificially high-redshift for the onset
of star formation.\\
For the test cases presented in this paper, the value adopted for the
high-density threshold is $135\,h^2\rm cm^{-3}$, well beyond the
isothermal peak of the gas. This allows a correct estimate
of the relevant timescales, as the gas spends most of the time in the
isothermal phase.
In addition, following the evolution of the gas to higher densities
allows higher resolution of, e.g., the morphology and disk
galaxy structure \cite[]{Saitoh_et_al_2008arXiv}, the clumpiness of
the gas, and the features of the interstellar or intergalactic  medium.
On the other hand, running high-density threshold simulations to
the present age ($z=0$) is computationally very challenging because of
the extremely short timescales involved in the calculations.
Only simulations performed with a low-density threshold are
currently run to $z=0$ and fine-tuned to reproduce the observed
low-redshift evolution of the star formation density.\\
We note that the mean-density region, because of
its small dimensions, lacks massive haloes.
In larger simulations, we expect to find rarer, larger, haloes, which can
grow faster  and host star formation in $\sim 10^5\,\rm M_\odot-10^6
\, M_\odot$ haloes.\\
We have performed high-resolution,
three-dimensional, N-body/SPH simulations including non-equilibrium
atomic and molecular chemistry, star formation prescriptions, and
feedback effects to investigate the onset of primordial star
formation.
We have studied how the primordial star formation rate changes
according to different gas-density threshold, cosmological
parameters, and simulation set-ups.
Our main findings are summarized in the following:

\begin{itemize}

\item
  The typical low-density thresholds (below $\sim 1\,\rm cm^{-3}$)
  are inadequate for describing star formation episodes in mean regions
  of the universe at high redshift. To correctly estimate the onset
  of star formation, high-density thresholds are necessary.

\item
  In rare, high-density peaks, the density can be higher than
  the usual low-density thresholds from very early times,
  therefore these prescriptions are not physically meaningful. Density
  thresholds lying beyond the isothermal peak (several particles per
  cm$^3$, in our case) are still required: they should satisfy the \cite{Bate_Burkert_1997} requirement ($N$ at least $\sim 2N_{neigh}$) and Eq. (\ref{crit_th}) of Section \ref{sect:dth}. However, as long as they are beyond the isothermal peak, given the faster evolution in the phase diagram of the cooling particles in dense environments, the very exact value is not crucial.

\item
  Different values of the threshold and the cosmological parameters
  can cause the onset of star formation at very different epochs:
  with a low-density threshold ($0.2\,h^2\rm\,cm^{-3}$), star formation
  starts at $z\sim 25-31$ (depending on the cosmology), while
  high-density threshold models ($135\,h^2\,\rm cm^{-3}$) predict a
  much later onset, at $z\sim 12-16$ (depending on the cosmology).

\item
  Performing primordial, rare, high-density region simulations within
  the high-density threshold model, we find that the local star
  formation can set in as early as $z\sim 48$. 

\end{itemize}

\noindent
We conclude by adding a few comments on the meaning of the term ``onset of star formation''.
From a purely physical point of view, the onset of star formation occurs when the proton-proton nuclear reactions ignite in a collapsed, dense core. Nowadays, in numerical simulations of cosmic structure evolution, this definition cannot be adopted, since it is not feasible to follow and resolve the behaviour of the gas on the very large range of scales involved. Therefore, the onset of star formation is meant as the attainment of a certain density threshold: once SPH particles reach it, they are assumed to be dense enough to collapse and to host star formation.
This threshold can be viewed as a point of no return, because it imposes a limit above which the natural gas evolution is strongly altered by star production and feedback effects.
The time when the threshold is reached and the time when the actual star formation takes place are typically not the same, but if the threshold is set appropriately, as discussed in the present work, they become very similar.
Broadly speaking, one could consider assigning a simple prescription to standard numerical simulations based on the typical delay time of gas in fall. In this way, the onset of star formation could be easily corrected without implementing high-density thresholds or molecular evolution.
In practice, this is not a trivial task: the time spent in the isothermal regime depends on numerous ``environmental'' factors, so different simulations with different initial conditions will be affected differently, according to their particular features. However, as an estimate, a typical free-fall time of $\sim 10^2~\rm Myr$ is the one we expect in correspondence of $\sim 10^{-1}~\rm cm^{-3}$ (conventional low-density thresholds).\\
Throughout this paper, we have considered the onset of star formation to be the attainment of the density threshold, provided that the Jeans mass is resolved by a ``large'' number of SPH particles and the isothermal peak in the phase diagram is resolved (and we have seen in Section \ref{sect:dth} and \ref{sect:results} that the latter two conditions are strongly related).

\begin{acknowledgements}
We acknowledge useful discussions with James Bolton, Massimo Ricotti, Cecilia Scannapieco, Volker Springel, Romain Teyssier, Michele Trenti, Simon D.~M.~White and John Wise.
NY thanks financial support from Grants-in-Aid for Young Scientists S from JSPS (20674003).
We also acknowledge the anonymous referee for stimulating comments on the paper.\\
The simulations were performed using the machines of the
Max Planck Society computing center, Garching
(Rechenzentrum-Garching) and of the Max-Planck-Institut f\"ur
Astrophysik.\\
For the bibliografic research we have made use of the tools offered by
the NASA Astrophysics Data System and by the JSTOR Archive.

\end{acknowledgements}

\bibliographystyle{aa}
\bibliography{paper.bib}


\end{document}